\documentstyle[onecolumn]{mn}

\title{ On the particle acceleration near the light surface of radio pulsars}

\author[V.S.~Beskin and R.R.~Rafikov]
       {V.S.~Beskin$^{1,2}$ and R.R.~Rafikov$^3$ \\
           $^1$ National Astronomical Observatory, Osawa 2--21--1,
                Mitaka, Tokyo 181--8588, Japan \\
           $^2$ P.N.Lebedev Physical Institute, Leninsky prosp., 53,
Moscow, 117924, Russia\\
           $^3$ Princeton University Observatory, Princeton, NJ, 08544, USA}
\date{Accepted 1999  .
     Received  1999 ;
     in original form 1999}

\pagerange{\pageref{firstpage}--\pageref{lastpage}}

\pubyear{1999}

\begin{document}

\maketitle

\label{firstpage}

\begin{abstract}

The two--fluid effects on the radial outflow of relativistic
electron--positron plasma are considered. It is shown that for
large enough Michel (1969) magnetization parameter $\sigma \gg 1$
and multiplication parameter $\lambda=n/n_{GJ} \gg 1$ one--fluid
MHD approximation remains correct in the whole region
$|{\bmath E}| < |{\bmath B}|$. In the case when the
longitudinal electric current is smaller than the Goldreich--Julian
one, 
the acceleration of
particles near the light surface $|{\bmath E}| = |{\bmath B}|$
is determined. It is shown that, as in the previously considered
(Beskin Gurevich \& Istomin 1983) cylindrical geometry,
almost all electromagnetic energy is transformed into the energy
of particles in the narrow boundary layer
$\Delta\varpi/\varpi \sim \lambda^{-1}$.

\end{abstract}

\begin{keywords}
two--fluid relativistic MHD:radio pulsars---particle acceleration
\end{keywords}

\section{Introduction}

Despite the fact that the structure of the magnetosphere of radio
pulsars remains
one of the fundamental astrophysical problems, the common view on the key
theoretical question -- what is the physical nature of the neutron star
braking -- is absent (Michel 1991, Beskin Gurevich \& Istomin 1993,
Mestel 1999). Nevertheless, very extensive theoretical studies in the
seventies and the eighties allowed to obtain some model-independent
results. One of them is the absence of magnetodipole energy loss.
This result was first obtained theoretically (Henriksen \& Norton
1975, Beskin et al 1983). It was shown that the electric charges
filling the magnetosphere screen fully the magnetodipole radiation
of a neutron star for an arbitrary inclination angle $\chi$ between
the rotational and magnetic axes if there are no longitudinal currents
flowing in the magnetosphere. Later this result was also confirmed
by observations. The direct detections of the interaction
of the pulsar wind with a  companion star in close binaries (see e.g.
Djorgovsky \& Evans 1988, Kulkarni \& Hester 1988) have shown that
it is impossible to explain the heating of the companion by
a low--frequency magnetodipole wave.

On the other hand, the detailed mechanism of particle acceleration
remains unclear. Indeed, a very high magnetization parameter $\sigma$
(Michel 1969) in the pulsar magnetosphere
demonstrates that within the light cylinder $r < R_{\rm L} = c/\Omega$
the main part of the energy is transported
by the Poynting flux. It means that
the additional mechanism of particle acceleration must work
in the vicinity of the light cylinder.
It is necessary to stress that 
an effective particle acceleration can only take place for small enough
longitudinal electric currents $I < I_{GJ}$ when the plasma has no
possibility to pass smoothly through the fast magnetosonic surface
and when the light surface $|{\bmath E}| = |{\bmath B}|$ is located at a
finite distance.
As to the case of the large longitudinal currents $I > I_{GJ}$,
both analytical (Tomimatsu 1994, Begelman \& Li 1994, Beskin et al 1998)
and numerical (Bogovalov 1997)
considerations demonstrate that the
 acceleration becomes ineffective
outside the fast magnetosonic surface, and the particle-to-Poynting
flux ratio remains small: $\sim \sigma^{-2/3}$ (Michel 1969, Okamoto 1978).

The acceleration of an electron--positron plasma near the light surface
was considered by Beskin Gurevich and Istomin (1983) in the simple
$1D$ cylindrical geometry for $I \ll I_{GJ}$.
It was shown that in a narrow boundary layer
$\Delta\varpi/\varpi \sim 1/\lambda$ almost all electromagnetic energy
is actually converted to the particles energy. Nevertheless, cylindrical
geometry does not provide the complete  picture of particle acceleration.
In particular, it was impossible to include self--consistently
the disturbance of a poloidal magnetic field and an electric potential,
the later playing the main role in the problem of the plasma acceleration
(for more details see e.g. Mestel \& Shibata 1994).
Hence, a more careful $2D$ consideration is necessary.

In Sect. 2 we formulate a complete system of $2D$ two--fluid MHD equations
describing the electron--positron outflow from a magnetized body with
a monopole magnetic field. The presence of an exact analytical force--free
solution (Michel 1973) allows us to linearize this system
which results in the existence of invariants (energy and angular momentum)
along unperturbed monopole field lines similar to the ideal one--fluid
MHD flow.
In Sect. 3 it is  shown that for $\sigma \gg 1$ and $\lambda \gg 1$
($\lambda=n/n_{GJ}$ is the multiplication factor)
the one--fluid MHD approximation remains true in the entire region within the
light surface. Finally, in Sect. 4 the acceleration of
particles near the light surface $|{\bmath E}| = |{\bmath B}|$
is considered. It is shown that, as in the case of cylindrical
geometry, in a narrow boundary layer
$\Delta\varpi/\varpi \sim \lambda^{-1}$ almost all the electromagnetic
energy is converted into the energy of particles.

\section{Basic Equations}

Let us consider a stationary axisymmetric outflow of a two--component
plasma in the vicinity of an active object with a monopole magnetic field.
It is necessary to stress that, of course, the monopole magnetic field
is a rather crude approximation for a pulsar magnetosphere. Nevertheless,
even for a dipole magnetic field near the origin, at large distances
$r \gg R_{\rm L}$ in the wind zone the magnetic field can have a monopole--like
structure. For this reason the disturbance of a monopole magnetic field
can give us an important information concerning particle acceleration
far from the neutron star.

The structure of the flow is described by the set of
Maxwell`s equations and the equations of motion
\begin{eqnarray}
\nabla{\bmath E}=4\pi \rho_e,~~~~~
\nabla\times{\bmath E} =0,\nonumber \\
\nabla{\bmath B}=0,~~~~~
\nabla\times{\bmath B} =\frac{4 \pi}{c}{\bmath j}, \label{cc1} \\
({\bmath v}^{\pm}\nabla){\bmath p}^{\pm}=
\pm e\left( {\bmath E}+ \frac{{\bmath v}^{\pm}}{c}\times{\bmath B}\right).
\label{eq:2}
\end{eqnarray}
Here $\bmath E$ and $\bmath B$ are the electric and magnetic
fields,  $\rho_e$ and $\bf j$ are the charge and
current densities, and ${\bmath v}^{\pm}$ and ${\bmath p}^{\pm}$  are
the speed and momentum of particles.
In the limit of infinite particle energy
\begin{equation}
\gamma=\infty, \quad
v_r^0=c, \quad
v_{\varphi}^0=0, \quad
v_{\theta}^0=0,
\end{equation}
and for charge and current density
\begin{equation}
\rho_e^0=\rho_{\rm s}\frac{R_{\rm s}^2}{r^2}\cos\theta, \quad
j_r = \rho_{\rm s}c\frac{R_{\rm s}^2}{r^2}\cos\theta,
\label{gj}
\end{equation}
the monopole poloidal magnetic field
\begin{equation}
B_r^0 = B_{\rm s}\frac{R_{\rm s}^2}{r^2}, \quad
B_{\theta}^0=0,~
\end{equation}
is the exact solution of Maxwell's equations. In this case
\begin{equation}
B_{\varphi}^0 = E_{\theta}^0 =-B_{\rm s}\frac{R_{\rm s}\Omega}{c}\frac{R_{\rm s}}{r}\sin\theta,
\quad E_r^0 = E_{\varphi}^0=0,
\end{equation}
which coincides with the well--known Michel (1973) solution.
Here $\gamma$ is the Lorentz-factor of particles,
$B_{\rm s}$ is the magnetic field on the surface of the sphere
$r = R_{\rm s}$, and $\rho_{\rm s} = $ const. As a result,
the angular velocity can be rewritten in a form
$\Omega = 2\pi c |\rho_{\rm s}|/B_{\rm s}$.
The limit $\gamma \rightarrow \infty$ just corresponds to zero particle
mass in the force--free approximation.

It is convenient to introduce the electric field potential $\Phi(r,\theta)$,
so that ${\bmath E}=-\nabla \Phi$ and
\begin{equation}
\Phi^0=-\frac{\Omega R_{\rm s}^2 B_{\rm s}}{c}\cos\theta,
\end{equation}
and the flux function $\Psi(r,\theta)$, so that
\begin{equation}
{\bmath B}_{\rm p}^0 = \frac{\nabla \Psi\times {\bmath e}_{\varphi}}
{2 \pi r\sin\theta},
\end{equation}
and $\Psi^0=2\pi B_{\rm s} R_{\rm s}^2(1-\cos\theta)$. Then one can seek the
first--order corrections for the case $v\ne c$ in the following form:
\begin{eqnarray}
n^{+} & = & \frac{\Omega B_{\rm s}}{2\pi c e}\frac{R_{\rm s}^2}{r^2}
\left[\lambda-\frac{1}{2}\cos\theta+\eta^{+}(r,\theta)\right],
\label{cc8}\\
n^{-} & = & \frac{\Omega B_{\rm s}}{2\pi c e}\frac{R_{\rm s}^2}{r^2}
\left[\lambda+\frac{1}{2}\cos\theta+\eta^{-}(r,\theta)\right],\\
\Phi(r,\theta) & = & \frac{\Omega R_{\rm s}^2
B_{\rm s}}{c}\left[-\cos\theta+\delta(r,\theta)\right],\\
\Psi(r,\theta) & = & 2\pi B_{\rm s} R_{\rm s}^2\left[1-\cos\theta+\varepsilon
f(r,\theta)\right],\\
v_r^{\pm} & = & c\left[1-\xi_r^{\pm}(r,\theta)\right],~~~v_{\theta}^{\pm}=
c\xi_{\theta}^{\pm}(r,\theta),~~~
v_{\varphi}^{\pm}=
c\xi_{\varphi}^{\pm}(r,\theta),\\
B_r & = & B_{\rm s}\frac{R_{\rm s}^2}{r^2}\left(
1+\frac{\varepsilon}{\sin\theta}
\frac{\partial f}{\partial \theta}\right),\\
B_{\theta} & = & -\varepsilon \frac{B_{\rm s} R_{\rm s}^2}{r \sin\theta}
\frac{\partial f}{\partial r},\\
B_{\varphi} & = & B_{\rm s}\frac{\Omega R_{\rm s}}{c}\frac{R_{\rm s}}{r}\left[-\sin\theta
-\zeta(r,\theta)\right],\\
E_r & = & -\frac{\Omega B_{\rm s} R_{\rm s}^2}{c}\frac{\partial \delta}{\partial r},\\
E_{\theta} & = & \frac{\Omega R_{\rm s}^2 B_{\rm s}}{c r}\left(
-\sin\theta - \frac{\partial \delta}{\partial \theta}\right).
\label{cc17}
\end{eqnarray}
Such a choice corresponds to a constant particle-to-magnetic flux ratio.
Here $\lambda \gg 1$ is the multiplication parameter
($\lambda=en_{\rm s}/|\rho_{\rm s}|$, where $n_{\rm s}$ is the concentration
of particles on the surface $r = R_{\rm s}$) which is $10^3 - 10^5$ for
radio pulsars. In what follows we consider for simplicity the case
$\lambda =$ const.

Substituting (\ref{cc8})--(\ref{cc17}) into equations 
(\ref{cc1})--(\ref{eq:2}), we obtain
to the first-order approximation the following system of equations:
\begin{eqnarray}
-\frac{1}{\sin\theta}\frac{\partial}
{\partial \theta}(\zeta\sin\theta)= 
2(\eta^+-\eta^-)
-2\left[\left(\lambda-\frac{1}{2}\cos\theta\right)\xi_r^+
-\left(\lambda+\frac{1}{2}\cos\theta\right)\xi_r^-\right],
\label{b1}\\
2(\eta^+-\eta^-)+\frac{\partial}{\partial r}\left(r^2
\frac{\partial \delta}{\partial r}\right)+
\frac{1}{\sin\theta}\frac{\partial}{\partial \theta}
\left(\sin\theta \frac{\partial\delta}{
\partial \theta}\right)=0,
\label{k1} \\
\frac{\partial\zeta}{\partial
r}=\frac{2}{r}
\left[\left(\lambda-\frac{1}{2}\cos\theta\right)\xi_{\theta}^+
-\left(\lambda+\frac{1}{2}\cos\theta\right)\xi_{\theta}^-\right],
\label{z1} \\
-\frac{\varepsilon}{\sin\theta}\frac{\partial^2 f}{\partial r^2}
-\frac{\varepsilon}{r^2}\frac{\partial}{\partial\theta}
\left(\frac{1}{\sin\theta}\frac{\partial f}
{\partial \theta}\right)= 
2\frac{\Omega}{r
c}\left[\left(\lambda-\frac{1}{2}\cos\theta\right)\xi_{\varphi}^+
-\left(\lambda+\frac{1}{2}\cos\theta\right)\xi_{\varphi}^-\right],
\label{n1}\\
\frac{\partial}{\partial r}\left(\xi_{\theta}^+\gamma^+\right)+
\frac{\xi_{\theta}^+\gamma^+}{r}=
4\lambda\sigma\left(
-\frac{1}{r}\frac{\partial\delta}{\partial\theta}+
\frac{\zeta}{r}-\frac{\sin\theta}{r}\xi_r^++
\frac{c}{\Omega r^2}\xi_{\varphi}^+\right),
\label{s1}\\
\frac{\partial}{\partial r}\left(\xi_{\theta}^-\gamma^-\right)+
\frac{\xi_{\theta}^-\gamma^-}{r}=
-4\lambda\sigma\left(
-\frac{1}{r}\frac{\partial\delta}{\partial\theta}+
\frac{\zeta}{r}-\frac{\sin\theta}{r}\xi_r^-+
\frac{c}{\Omega r^2}\xi_{\varphi}^-\right),
\label{z2}\\
\frac{\partial}{\partial r}\left(\gamma^+\right)=
4\lambda\sigma\left(
-\frac{\partial\delta}{\partial r}-
\frac{\sin\theta}{r}\xi_{\theta}^+\right),
\label{s2}\\
\frac{\partial}{\partial r}\left(\gamma^-\right)
=-4\lambda\sigma\left(
-\frac{\partial\delta}{\partial r}-
\frac{\sin\theta}{r}\xi_{\theta}^-\right),
\label{z3} \\
\frac{\partial}{\partial r}\left(\xi_{\varphi}^+\gamma^+\right)+
\frac{\xi_{\varphi}^+\gamma^+}{r}=
4\lambda\sigma\left(
-\varepsilon\frac{c}{\Omega r \sin\theta}\frac{\partial f}
{\partial r}-\frac{c}{\Omega r^2}\xi_{\theta}^+\right), \\
\frac{\partial}{\partial r}\left(\xi_{\varphi}^-\gamma^-\right)+
\frac{\xi_{\varphi}^-\gamma^-}{r}=
-4\lambda\sigma\left(
-\varepsilon\frac{c}{\Omega r \sin\theta}\frac{\partial f}
{\partial r}-\frac{c}{\Omega r^2}\xi_{\theta}^-\right).
\label{b2}
\end{eqnarray}
Here
\begin{equation}
\sigma=\frac{\Omega e B_{\rm s} R_{\rm s}^2}{4\lambda m c^3}
\end{equation}
is the Michel`s (1969) magnetization parameter,
$m$ is the electron mass,
and all deflecting functions are supposed to be $\ll 1$.
It is necessary to stress that for applications the magnetic field
$B_{\rm s}$ is to be taken near the light cylinder
$R_{\rm s} \approx R_{\rm L}$ because in the internal region
of the pulsar magnetosphere $B \propto r^{-3}$.
As it has already been mentioned, only outside
the light cylinder the poloidal magnetic field may have quasi
monopole structure. As a result,
\begin{equation}
\sigma=\frac{\Omega^2 e B_0 R^3}{4\lambda m c^4}
\approx 10^{4}B_{12}\lambda_{3}^{-1}P^{-2},
\end{equation}
where $B_0$ -- magnetic field on the neutron star surface $r = R$.
Hence, for ordinary pulsars ($P \sim 1$s, $B_0 \sim 10^{12}$G) we have
$\sigma \sim 10^{4} - 10^{5}$, and only for fast ones
($P \sim 0.1 - 0.01$s, $B_0 \sim 10^{13}$G) we have
$\sigma \sim 10^{6} - 10^{7}$.

Formally, this system of equations requires twelve boundary conditions.
We consider for simplicity the case $\Omega R/c \ll 1$ when the star
radius $R$ is much smaller than the light cylinder.
As a result, one writes down
the first six boundary conditions as
\begin{eqnarray}
\xi_{\theta}^{\pm}(R_{\rm s},\theta) & = & 0, \\
\xi_{\varphi}^{\pm}(R_{\rm s},\theta) & = & 0, \\
\gamma^{\pm}(R_{\rm s}, \theta) & = & \gamma_{\rm in},
\end{eqnarray}
i.e. $\xi_r^{\pm}(R_{\rm s},\theta) = 1/(2\gamma_{\rm in}^2)$.
According to all theories of particle generation near the neutron star
surface (Ruderman Sutherland 1975, Arons Scharlemann 1979),
$\gamma_{\rm in} \leq 10^2$ for secondary plasma. For this reason in what
follows we consider in more details the case
\begin{equation}
\gamma_{\rm in}^3 \ll \sigma, 
\label{sgm}
\end{equation}
when the additional acceleration of particles
inside the fast magnetosonic surface takes place
(see e.g. Beskin Kuznetsova Rafikov 1998).
It is this case that can be realized for fast pulsars. Moreover, it has
more general interest because the relation (\ref{sgm}) may be true also
for AGNs.
As to the case $\gamma_{\rm in}^3 \gg \sigma$ corresponding to ordinary
pulsars, the particle energy remains constant ($\gamma = \gamma_{\rm in}$)
at any way up to the fast magnetosonic surface (see Bogovalov 1997 for details).

Further, one can put
 \begin{eqnarray}
\delta(R_{\rm s},\theta)  & = & 0,
\label{bc1}\\
\varepsilon f(R_{\rm s},\theta) & = & 0,
\label{bc2}\\
\eta^+(R_{\rm s},\theta)-\eta^-(R_{\rm s},\theta) & = & 0.
\end{eqnarray}
These conditions result from the relation
$c{\bmath E}_{\rm s}
+ \Omega R_{\rm s}{\bmath e}_{\varphi} \times {\bmath B}_{\rm s} = 0$
corresponding rigid rotation and perfect conductivity of the surface
of a star. Finally, as will be shown in Sect. 3.2, the derivative
$\partial\delta/\partial r$ actually determines  the phase of plasma
oscillations only and plays no role in the global structure. Finally,
the determination of the electric current and, say, the derivative
$\partial f/\partial r$ depend on the problem under consideration.
Indeed, as is well--known, the cold one--fluid MHD outflow contains
two singular surfaces, Alfv\'enic and fast magnetosonic ones. It means that
for the transonic flow two latter functions are to be determined from
critical conditions (Heyvaerts 1996). In particular, the longitudinal
electric current within this approach is not a free parameter. On the
other hand, if the electric current is restricted by some physical
reason, the flow cannot pass smoothly through the fast magnetosonic surface.
In this case, which can be realized in the magnetosphere of radio
pulsars (Beskin et al 1983, Beskin \& Malyshkin 1998), near the
light surface $|{\bmath E}| = |{\bmath B}|$ an effective particle
acceleration may take place. Such an acceleration will be considered
in Sect. 4.

\section{The electron--positron outflow}

\subsection{The MHD Limit}

In the general case Eqns. (\ref{b1}) -- (\ref{b2}) have several
integrals. Firstly, Eqns.(\ref{z1}), (\ref{s2}), and (\ref{z3}) result in
\begin{equation}
\zeta-\frac{2}{\tan\theta}\delta
+\frac{(\lambda-1/2\cos\theta)\gamma^++(\lambda+1/2\cos\theta)\gamma^-}
{2\sigma\lambda\sin\theta}=
\frac{1}{\sigma\sin\theta}\gamma_{\rm in}+\frac{l(\theta)}{\sin\theta},
\label{1}
\end{equation}
where $l(\theta)$ describe the disturbance of the electric current
at the star surface by the equation
$I(R,\theta)=I_A\left[\sin^2\theta+l(\theta)\right]$.
Expression (\ref{1}) corresponds
to conservation of the total energy flux along a magnetic field line.
Furthermore, Eqns. (\ref{s2}) -- (\ref{b2}) together with the boundary
conditions (\ref{bc1}), (\ref{bc2}) result in
\begin{eqnarray}
\delta & = & \varepsilon f-\frac{1}{4\lambda\sigma}\gamma^+
\left(1-\frac{\Omega r\sin\theta}{c}\xi_{\varphi}^+\right)
+\frac{1}{4\lambda\sigma}\gamma_{\rm in};
\label{7}\\
\delta & = & \varepsilon f+\frac{1}{4\lambda\sigma}\gamma^-
\left(1-\frac{\Omega r\sin\theta}{c}\xi_{\varphi}^-\right)
-\frac{1}{4\lambda\sigma}\gamma_{\rm in}.
\label{8}
\end{eqnarray}
They correspond to conservation of the $z$--component of the angular momentum
for both types of particles. It is necessary to stress that the
complete nonlinearized
system of equations contains no such simple invariants.

As $\sigma\lambda \gg 1$, we can neglect in Eqns. (\ref{s1})--(\ref{b2})
their left-hand sides. In this approximation we have $\xi^+ = \xi^-$
i.e. $\gamma^- = \gamma^+ = \gamma$, so that
\begin{eqnarray}
-\frac{1}{r}\frac{\partial\delta}{\partial\theta}+
\frac{\zeta}{r}-\frac{\sin\theta}{r}\xi_r+
\frac{c}{\Omega r^2}\xi_{\varphi}=0,
\label{6} \\
\varepsilon\frac{c}{\Omega r \sin\theta}\frac{\partial f}
{\partial r}+ \frac{c}{\Omega r^2}\xi_{\theta}=0,
\label{6a}
\end{eqnarray}
and
\begin{equation}
\gamma\left(1-\frac{\Omega r\sin\theta}{c}\xi_{\varphi}\right)=\gamma_{\rm in}.
\label{5}
\end{equation}
Hence, within this approximation
\begin{eqnarray}
\delta & = & \varepsilon f, \\
\zeta & = & \frac{2}{\tan\theta}\varepsilon f+\frac{l(\theta)}{\sin\theta}-
\frac{1}{\sigma\sin\theta}(\gamma-\gamma_{\rm in}).
\end{eqnarray}
Substituting these expressions into (\ref{6}) and using Eqns.
(\ref{b1})--(\ref{n1}), we obtain the following equation describing
the disturbance of the magnetic surfaces
\begin{eqnarray}
\varepsilon (1-x^{2}\sin^{2}\theta)
\frac{\partial^{2}f}{\partial x^2}
+\varepsilon(1-x^{2}\sin^{2}\theta)
\frac{\sin\theta}{x^2}\frac{\partial}{\partial\theta}
\left(\frac{1}{\sin\theta}\frac{\partial f}{\partial\theta}\right)
- 2\varepsilon x\sin^{2}\theta\frac{\partial f}{\partial x}
-2\varepsilon\sin\theta\cos\theta\frac{\partial f}{\partial\theta}
+2\varepsilon(3\cos^{2}\theta-1)f \\
+\frac{1}{\sin\theta}\frac{{\rm d}}{{\rm d}\theta}(l\sin^2\theta)
-2\frac{\cos\theta}{\sigma}\left(\gamma-\gamma_{\rm in}\right)
-\frac{\sin\theta}{\sigma}\frac{\partial\gamma}{\partial\theta}
-2\lambda\sin^2\theta(\xi_{r}^+-\xi_{r}^-)
+\frac{2\lambda}{x}\sin\theta(\xi_{\varphi}^+-\xi_{\varphi}^-)=0,
\nonumber
\label{mnq}
\end{eqnarray}
where $x=\Omega r/c$. One can see
that it actually coincides with the one--fluid MHD
Eqns.(32), (52) from Beskin et al (1998), but contains the two last
additional nonhydrodynamical terms. Nevertheless, as will be shown
in the next subsection, at small distances $r \ll r_{\rm F}$
where $r_{\rm F}$ is the radius of the fast magnetosonic surface
we have
\begin{equation}
-\lambda\sin^2\theta(\xi_{r}^+-\xi_{r}^-)
+\frac{\lambda}{x}\sin\theta(\xi_{\varphi}^+-\xi_{\varphi}^-) \approx 0,
\label{add}
\end{equation}
so actually there is perfect agreement with the MHD approximation
\begin{eqnarray}
\varepsilon (1-x^{2}\sin^{2}\theta)
\frac{\partial^{2}f}{\partial x^2}
+\varepsilon(1-x^{2}\sin^{2}\theta)
\frac{\sin\theta}{x^2}\frac{\partial}{\partial\theta}
\left(\frac{1}{\sin\theta}\frac{\partial f}{\partial\theta}\right)
- 2\varepsilon x\sin^{2}\theta\frac{\partial f}{\partial x}
-2\varepsilon\sin\theta\cos\theta\frac{\partial f}{\partial\theta} 
\label{mn1}\\
+2\varepsilon(3\cos^{2}\theta-1)f
+\frac{1}{\sin\theta}\frac{{\rm d}}{{\rm d}\theta}(l\sin^2\theta)
-2\frac{\cos\theta}{\sigma}\left(\gamma-\gamma_{\rm in}\right)
-\frac{\sin\theta}{\sigma}\frac{\partial\gamma}{\partial\theta}
= 0.
\nonumber
\end{eqnarray}

As was shown earlier (Beskin et al 1998), to pass through
the fast magnetosonic surface it's necessary to have
\begin{equation}
|l| < \sigma^{-4/3}.
\label{nn}
\end{equation}
Hence, within the fast magnetosonic surface $r \ll r_{\rm F}$
one can neglect terms containing $\delta = \varepsilon f$ and $\zeta$.
Then, relations (\ref{6}) and (\ref{6a}) result in
\begin{eqnarray}
\gamma(1-x\sin\theta\xi_{\varphi}) & = & \gamma_{\rm in}, \\
\xi_r & = & \frac{\xi_{\varphi}}{x\sin\theta}, \\
\xi_{\theta}& = & 0.
\end{eqnarray}
Finally, using the definition
\begin{equation}
\gamma^2=\frac{1}{2\xi_r-\xi_{\varphi}^2},
\label{3}
\end{equation}
we obtain for $\sigma \gg \gamma_{\rm in}^3$ for $r \ll r_{\rm F}$
\begin{eqnarray}
\gamma^2 & = & \gamma_{\rm in}^2+x^2\sin^2\theta,
\label{v0}\\
\xi_{\varphi} & = &
\frac{\sqrt{\gamma_{\rm in}^2+x^2\sin^2\theta}-\gamma_{\rm in}}
{x\sin\theta\sqrt{\gamma_{\rm in}^2+x^2\sin^2\theta}}
\sim \frac{1}{x\sin\theta},
\label{v1} \\
\xi_r & = &
\frac{\sqrt{\gamma_{\rm in}^2+x^2\sin^2\theta}-\gamma_{\rm in}}
{x^2\sin^2\theta\sqrt{\gamma_{\rm in}^2+x^2\sin^2\theta}}
\sim \frac{1}{x^2\sin^2\theta},
\label{v2}
\end{eqnarray}
in full agreement with the MHD results.

Next, to determine the position of the fast magnetosonic surface $r_{\rm F}$,
one can analyze the algebraic equations (\ref{1}) and (\ref{6}) which
give
\begin{equation}
-\frac{\partial\delta}{\partial\theta}+\frac{2}{\tan\theta}\delta
-\frac{1}{\sigma\sin\theta}\gamma-\sin\theta\xi_r
+\frac{1}{x}\xi_{\varphi}=0.
\end{equation}
Using now expressions (\ref{5}) and (\ref{3}), one can find
\begin{equation}
2\gamma^3-2\sigma\left[K+\frac{1}{2x^2}\right]\gamma^2
+\sigma\sin^2\theta = 0,
\label{9}
\end{equation}
where
\begin{equation}
K(r,\theta)=2\cos\theta\delta-\sin\theta\frac{\partial\delta}{\partial\theta}.
\end{equation}

Equation (\ref{9}) allows us to determine the position of the fast magnetosonic
surface and the energy of particles. Indeed, determining the derivative
$r\partial\gamma/\partial r$, one can obtain
\begin{equation}
r\frac{\partial\gamma}{\partial r}=
\frac{\gamma\sigma\left(r\partial K/\partial r-x^{-2}\right)}
{3\gamma-\sigma\left(2K+x^{-2}\right)}.
\end{equation}
As the fast magnetosonic surface is the $X$--point, both the nominator and
denominator are to be equal to zero here.
As a result, evaluating $r\partial K/\partial r$ as $K$, we obtain
\begin{eqnarray}
\delta & \sim & \sigma^{-2/3}; \\
r_{\rm F} & \sim & \sigma^{1/3}R_{\rm L};
\label{51}\\
\gamma(r_{\rm F}) & = & \sigma^{1/3}\sin^{2/3}\theta,
\end{eqnarray}
where the last expression is exact. These equations
 coincide with those obtained
within the MHD consideration.  It is the self--consistent analysis when
$\delta=\varepsilon f$, and hence $K$ depends on the radius $r$ that results in
the finite value for the fast magnetosonic radius $r_{\rm F}$.  On the other hand, in
a given monopole magnetic field, when $\varepsilon f$
does not depend on the radius,
the critical conditions result in $r_{\rm F}\rightarrow \infty$ for a cold outflow
(Michel, 1969, Li et al 1992).

Near the fast magnetosonic surface $r \sim \sigma^{1/3}R_{\rm L}$ the MHD
solution gives
\begin{eqnarray}
\gamma & \sim & \sigma^{1/3}, \\
\varepsilon f & \sim & \sigma^{-2/3}.
\end{eqnarray}
Hence, Eqns. (\ref{3}), (\ref{v1}), and
 (\ref{v2}) result in
\begin{eqnarray}
\xi_r & \sim & \sigma^{-2/3}, \\
\xi_{\theta} & \sim & \sigma^{-2/3}, \\
\xi_{\varphi} & \sim & \sigma^{-1/3}.
\end{eqnarray}
As we see, the $\theta$--component of the velocity plays no role
in the determination of the $\gamma$.

However, analyzing the
left-hand sides of the Eqns. (\ref{s1})--(\ref{b2})
one can evaluate the additional (nonhydrodynamic) variations of the velocity
components
\begin{eqnarray}
\Delta\xi_r^{\pm} & \sim & \lambda^{-1}\sigma^{-4/3}, \\
\Delta\xi_{\theta}^{\pm} & \sim & \lambda^{-1}\sigma^{-2/3}, \\
\Delta\xi_{\varphi}^{\pm} & \sim & \lambda^{-1}\sigma^{-1}.
\end{eqnarray}
Hence, for nonhydrodynamic velocities $\Delta\xi_r^{\pm} \ll \xi_r$ and
$\Delta\xi_{\varphi}^{\pm} \ll \xi_{\varphi}$ to be small,
it is necessary to have a large magnetization parameter $\sigma
\gg 1$ only. On the other hand,
$\Delta\xi_{\theta}^{\pm}/\xi_{\theta}\sim \lambda^{-1}$.
In other words, for a highly magnetized plasma $\sigma \gg 1$
even outside the fast magnetosonic surface
the velocity components (and, hence, the
particle energy) can be considered hydrodynamically.
The difference $\sim \lambda^{-1}$ appears in the $\theta$ component
only, but it does not affect the particle energy. Finally,
one can obtain from (\ref{7}), (\ref{8}) that
\begin{equation}
\frac{\delta-\varepsilon f}{\varepsilon f} \sim \lambda^{-2}\sigma^{-2/3}.
\end{equation}
To put it differently, at large distances the nonhydrodynamical terms are much
smaller than hydrodynamical ones.

As a result, at large distances where, according to (\ref{7})--(\ref{8}),
one can neglect the toroidal component $\xi_{\varphi}$, we obtain
\begin{eqnarray}
\delta & = & \varepsilon f, \\
\zeta & = & \frac{2}{\tan\theta}\delta
-\sigma^{-1}\frac{1}{\sin\theta}\gamma.
\label{z1a}
\end{eqnarray}
On the other hand, Eqn. (\ref{s1}) gives
\begin{equation}
\zeta=\frac{\partial\delta}{\partial\theta}+\sin\theta\xi_r.
\end{equation}
Together with (\ref{z1}) one can obtain for $r \gg r_{\rm F}$
\begin{equation}
\gamma=\sigma\left(2\cos\theta\varepsilon f
-\varepsilon\sin\theta\frac{\partial
f}{\partial\theta}\right),
\label{k2}
\end{equation}
which coincides with the MHD solution. Finally,
using Eqns. (\ref{b1}), (\ref{k1}), and neglecting the nonhydrodynamic
term $4\lambda(\xi_r^+-\xi_r^-)$, one can find
\begin{equation}
\varepsilon\frac{\partial}{\partial r}\left(r^2\frac{\partial f}
{\partial r}\right)
-4\cos\theta\xi_r
-\sin\theta\frac{\partial}{\partial\theta}\xi_r+
\frac{1}{x\sin\theta}\frac{\partial}{\partial\theta}(\xi_{\varphi}\sin\theta)
=0.
\label{pp}
\end{equation}
Together with (\ref{k2}) this equation in the limit $r \gg r_{\rm F}$
coincides with the asymptotic version of the trans--field equation
(Tomimatsu 1994, Beskin et al 1998)
\begin{equation}
\varepsilon\frac{\partial^2 f}{\partial r^2}+
2\varepsilon r\frac{\partial f}{\partial r}
-\sin\theta\frac{D+1}{D}\frac{\partial g}{\partial\theta} = 0,
\end{equation}
where $g(\theta)=K(\theta)/\sin^2\theta$, and
\begin{equation}
D+1 = \frac{1}{\sigma^2\sin^4\theta}g^{-3}(\theta) \ll 1.
\end{equation}
In this limit, none of the terms containing $\xi_r^{\pm}$ and $\xi_{\varphi}$
plays role in the asymptotic trans--field equation. Hence, it
is not necessary to consider the effect of the nonhydrodynamical term
$4\lambda(\xi_r^+-\xi_r^-)$ either.

\subsection{Plasma Oscillations}

In the intermediate region $r \ll r_{\rm F}$
Eqn. (\ref{pp}) cannot be used. The point is that in the limit
$\lambda \gg 1$ the important role in Eqns. (\ref{b1}) and (\ref{n1})
is played by the nonhydrodynamic terms (\ref{add})
corresponding to different velocities of two
components. As a result, the full version of Eqn. (\ref{pp}) has the form
\begin{equation}
\frac{\partial}{\partial r}\left(r^2\frac{\partial\delta}
{\partial r}\right)
-4\cos\theta\xi_r
-\sin\theta\frac{\partial\xi_r}{\partial\theta}+
\frac{1}{x\sin\theta}\frac{\partial}{\partial\theta}(\xi_{\varphi}\sin\theta)
+2\lambda(\xi_r^+-\xi_r^-)=0.
\label{pk}
\end{equation}
Indeed, one can see from equations (\ref{b1}) and (\ref{k1}) that near
the origin $x=R_{\rm s}$ in the case $\gamma_{\rm in}^+=\gamma_{\rm in}^-$
(and for the small variation of the current $\zeta \sim \sigma^{-4/3}$
which is necessary, as was already stressed, to pass through a fast
magnetosonic surface) the  density variation on the surface is large enough:
$(\eta^+-\eta^-) \sim \gamma_{\rm in}^{-2} \gg \zeta$. Hence, the derivative
$\partial^2\delta/\partial r^2$ here is of the order of $\gamma_{\rm in}^{-2}$.
On the other hand, according to (\ref{n1}), the derivative
$\varepsilon\partial^2 f/\partial r^2$ is $x^2$ times smaller.
This means that in the two--component system the longitudinal electric field
is to appear resulting in a redistribution of the particle energy.
Clearly, such a redistribution is impossible for the charge--separated outflow.
In other words, for a finite particle energy a one--component plasma cannot
maintain simultaneously both the Goldreich charge and Goldreich
current density (\ref{gj}).
In a two--component system with $\lambda \gg 1$ it can be realized by a small
redistribution of particle energy (Ruderman \& Sutherland 1975,
Arons \& Scharlemann 1989).

For simplicity, let us consider only  small distances $x \ll 1$.
In this case one can neglect the changes of the magnetic surfaces.
Using now (\ref{s2}) and (\ref{z3}), we have
\begin{eqnarray}
\gamma^+ & = & \gamma_{\rm in}-4\lambda\sigma\delta; \\
\gamma^- & = & \gamma_{\rm in}+4\lambda\sigma\delta.
\end{eqnarray}
Finally, taking into account that $\xi_{\theta}$ and $\xi_{\varphi}$ are
small here, one can obtain from (\ref{k1}) 
\begin{equation}
r^2\frac{\partial^2\delta}{\partial r^2}+2r\frac{\partial\delta}{\partial r}
+\frac{1}{\sin\theta}\frac{\partial}{\partial\theta}\left(\sin\theta
\frac{\partial\delta}{\partial\theta}\right)
+A\delta=\frac{\cos\theta}{\gamma_{\rm in}^2},
\label{p3}
\end{equation}
where
\begin{equation}
A=16\frac{\lambda^2\sigma}{\gamma_{\rm in}^3} \gg 1.
\end{equation}
Eqn. (\ref{p3}) has a solution
\begin{equation}
\delta=
\delta_0+r^{-1/2}
\left[C_1\sin(\mu\ln r)+C_2\cos(\mu\ln r)\right]\cos\theta,
\label{p4}
\end{equation}
where
\begin{equation}
\delta_0 \approx
\frac{\gamma_{\rm in}\cos\theta}{16\lambda^2\sigma},
\end{equation}
and $\mu\approx \sqrt{A}$. As we see, Eqn. (\ref{p4})
describes plasma oscillations similar to those considered by
Shibata (1997) for charge--separated flow. The decrease of
oscillations results from a more accurate consideration of
the Laplace operator in a 3D space.

One can easily check that the additional potential $\delta_0$ is small,
and it is not necessary to add it in (\ref{1}) and (\ref{7})--(\ref{8}).
Moreover, the nonhydrodynamic disturbance $\Delta\xi_r$ (as well as
$\Delta\gamma$) is also small, $\Delta\xi_r/\xi_r \approx \lambda^{-1}$.
Hence, as was already stressed, the boundary condition $\partial\delta/\partial
r$ (determining together with (\ref{bc1}) the coefficients $C_1$ and $C_2$)
does not affect the general structure of the flow.
On the other hand, the presence of an additional electric potential $\delta_0$
results in a full compensation of the last term in (\ref{b1})
\begin{equation}
2\lambda(\xi_r^+-\xi_r^-)-\cos\theta\xi_r \approx 0.
\label{u1}
\end{equation}
Next, as $\varepsilon f \ll \sigma^{-2/3}$ for $r \ll r_{\rm F}$, a similar
expression can be written for the $\varphi$--components as well
\begin{equation}
2\lambda(\xi_{\varphi}^+-\xi_{\varphi}^-)-\cos\theta\xi_{\varphi} \approx 0.
\label{u2}
\end{equation}
Expressions (\ref{u1}) -- (\ref{u2}) must hold for the whole region $r < r_{\rm F}$.
In this case, the final version
of Eqn. (\ref{pk}) in the internal region $r \ll r_{\rm F}$ can be
rewritten as
\begin{equation}
\frac{\partial}{\partial r}\left(r^2\frac{\partial\delta}
{\partial r}\right)
-2\cos\theta\xi_r
-\sin\theta\frac{\partial}{\partial\theta}\xi_r+
\frac{1}{x\sin\theta}\frac{\partial}{\partial\theta}(\xi_{\varphi}\sin\theta)
=0.
\label{ps}
\end{equation}

As $\delta \sim \varepsilon f \ll \sigma^{-2/3}$ for $r \ll r_{\rm F}$,
and $\xi_r \sim \gamma_0^{-2} \gg \delta$, the first
term in (\ref{ps}) can be omitted. As a result, the solution
of Eqn. (\ref{ps}) coincides exactly with the MHD expression, i.e.
$\gamma^2=\gamma_{\rm in}^2+x^2\sin^2\theta$ (\ref{v0}).
Finally, using (\ref{u1}), (\ref{u2}), and (\ref{v1})--(\ref{v2}),
one can easily check that the nonhydrodynamical terms (\ref{add}) in
the trans--field equation (\ref{mn1}) do actually vanish.

\section{The Boundary Layer}

Let us now consider the case when the longitudinal electric current
$I(R,\theta)$ in the magnetosphere of radio pulsars is too small
(i.e. the disturbance $l(\theta)$ is too large) for the flow to
pass smoothly through the fast magnetosonic surface. First of all,
it can be realized when the electric current is much smaller
than the Goldreich one. This possibility was already discussed within
the Ruderman--Sutherland model of the internal gap (Beskin et al 1983,
Beskin \& Malyshkin 1998). But it may take place in the Arons model
(Arons \& Scharlemann 1979) as well. Indeed, within this model the
electric current is determined by the gap structure. Hence, in general
case this current does not correspond to the critical condition at the
fast magnetosonic surface. In particular, it may be smaller than the
critical current (of course, the separate consideration is necessary to
check this statement).

For simplicity let us consider the case $l(\theta)=h\sin^2\theta$.
Neglecting now the last terms $\propto \sigma^{-1}$ in the trans--field
equation (\ref{mn1}), we obtain
\begin{eqnarray}
\varepsilon (1-x^{2}\sin^{2}\theta)
\frac{\partial^{2}f}{\partial x^2}
+\varepsilon(1-x^{2}\sin^{2}\theta)
\frac{\sin\theta}{x^2}\frac{\partial}{\partial\theta}
\left(\frac{1}{\sin\theta}\frac{\partial f}{\partial\theta}\right)
\nonumber \\
- 2\varepsilon x\sin^{2}\theta\frac{\partial f}{\partial x}
-2\varepsilon\sin\theta\cos\theta\frac{\partial f}{\partial\theta}
+2\varepsilon(3\cos^{2}\theta-1)f
+4h\sin^2\theta\cos\theta = 0,
\end{eqnarray}
which actually coincides with the force--free equation (Beskin et al 1998).
This equation has an exact analytical solution
\begin{equation}
\varepsilon f=hx^2\sin^2\theta\cos\theta.
\label{qa}
\end{equation}
For $h < 0$ (when the electric current is smaller than the
Goldreich one) this solution results in the appearance of the light
surface $|{\bmath E}|=|{\bmath B}|$ at the finite distance
\begin{equation}
\varpi_c = \frac{R_{\rm L}}{(2|h|)^{1/4}}.
\label{qq}
\end{equation}
As we see, for $l(\theta)=h\sin^2\theta$ this surface has the  form
of a cylinder. It is important that the disturbance of magnetic
surfaces $\varepsilon f \sim
(|h|)^{1/2}$ remains small here.

Comparing now the position of the light surface (\ref{qq}) with that of the
fast magnetosonic surface (\ref{51}), one can find that the light surface
is located inside the fast magnetosonic one if
\begin{equation}
\sigma^{-4/3} \ll |h| \ll 1,
\label{qs}
\end{equation}
which is opposite to (\ref{nn}).
One can check that the condition (\ref{qs}) just allows to neglect
the non force--free term in Eqn. (\ref{mn1}).

\begin{figure}
\vspace{10.cm}
\includegraphics{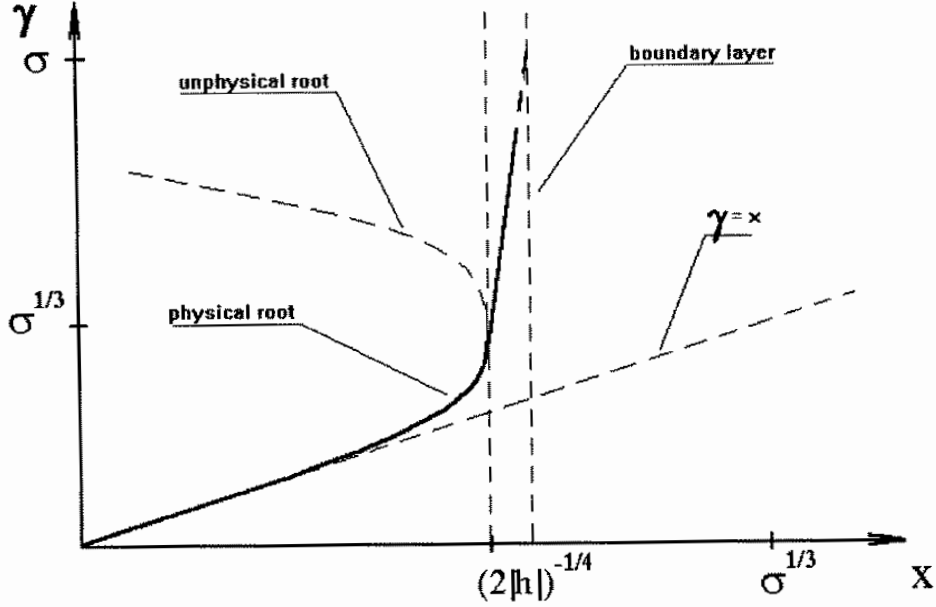}
\caption{
The behavior of the Lorentz factor in the case $\sigma^{-4/3} \ll |h| \ll 1$.
One can see that the one--fluid MHD solution (\ref{fig}) exists for
$\gamma < \sigma^{1/3}$ only. But in the two--fluid approximation
in the narrow layer $\Delta\varpi = \varpi_c/\lambda$
the particle energy increases up to the value $\sim \sigma$
corresponding to the full conversion of the electromagnetic
energy to the energy of particles.}
\label{figure}
\end{figure}

Using now the  solution (\ref{qa}) and the MHD condition
$\delta=\varepsilon f$, one can find from (\ref{9})
\begin{equation}
2\gamma^3-2\sigma\left(hx^2\sin^4\theta+\frac{1}{2x^2}\right)\gamma^2
+\sigma\sin^2\theta=0.
\end{equation}
This equation shows that near the force--free boundary
$x_{ff} = (2|h|)^{-1/4}$ (\ref{qq})
\begin{equation}
\gamma = \sigma^{1/3}\sin^{2/3}\theta -
\frac{2|2h|^{3/8}}{\sqrt{3}}\sigma^{1/3}\sin^{4/3}\theta
\sqrt{x_0-x\sin\theta},
\label{fig}
\end{equation}
where
\begin{equation}
x_0 = \frac{1}{(2|h|)^{1/4}}\left[1-\frac{3}{4(2|h|)^{1/2}}
\frac{1}{(\sigma\sin^2\theta)^{2/3}}\right],
\label{qw}
\end{equation}
(see Fig. \ref{figure}).
Hence, the real solution is absent for $x\sin\theta > x_0$.
Here $\gamma(x_0)=\sigma^{1/3}\sin^{2/3}\theta$,
the condition $\sigma^{-4/3} \ll |h|$ resulting in $\varpi_c < r_{\rm F}$,
and the last term in (\ref{qw}) being small.

Although the energy of particles at the limiting point is finite,
the derivative $d\gamma/d\varpi$ moves to infinity.
Hence, near the light surface the left-hand sides in the Eqns.
(\ref{s1}) -- (\ref{b2}) are to be taken into consideration.
Since in our case the light surface has the  form of a cylinder,
one can move to derivatives perpendicular to the boundary layer
only by
\begin{eqnarray}
\partial/\partial r \rightarrow \sin\theta\partial/\partial\varpi; \\
\partial/\partial\theta \rightarrow \varpi\cos\theta\partial/\partial\varpi.
\end{eqnarray}
As a result, $\zeta$ can be eliminated from (\ref{b1}) and
(\ref{z1}).  Together with (\ref{k1}) they give the equation for
$\delta$ (see (\ref{p6})).  Next, the invariants (\ref{7}
) and
(\ref{8}) can be used to define $\xi_{\varphi}^{\pm}$:
\begin{eqnarray}
\xi_{\varphi}^+ & = & \frac{1}{x\sin\theta}
\left[1+\frac{4\lambda\sigma(\delta-\varepsilon f)}{\gamma^+}\right];
\label{p1}\\
\xi_{\varphi}^- & = & \frac{1}{x\sin\theta}
\left[1-\frac{4\lambda\sigma(\delta-\varepsilon f)}{\gamma^-}\right].
\label{p2}
\end{eqnarray}
Furthermore, one can define
\begin{eqnarray}
2\xi_r^+ & = & \frac{1}{(\gamma^+)^2}+(\xi_{\varphi}^+)^2+(\xi_{\theta}^+)^2;
\label{p3a}\\
2\xi_r^- & = & \frac{1}{(\gamma^-)^2}+(\xi_{\varphi}^-)^2+(\xi_{\theta}^-)^2.
\label{p4a}
\end{eqnarray}
As to the energy integral (\ref{1}), it determines the variation of
the current $\zeta$. Now it can be rewritten as
\begin{equation}
\zeta=\frac{2}{\tan\theta}\delta
-\frac{(\gamma^++\gamma^-)}{2\sigma\sin\theta}.
\label{p5}
\end{equation}
Finally, Eqns. (\ref{b1}) -- (\ref{b2}) look like
\begin{eqnarray}
\varpi_c^2\frac{d^2\delta}{d\varpi^2}=
2\sin\theta\cos\theta
\left[\left(\lambda-\frac{1}{2}\cos\theta\right)\xi_{\theta}^+
-\left(\lambda+\frac{1}{2}\cos\theta\right)\xi_{\theta}^-\right]
\nonumber \\
-2\sin^2\theta\left[\left(\lambda-\frac{1}{2}\cos\theta\right)\xi_r^+
-\left(\lambda+\frac{1}{2}\cos\theta\right)\xi_r^-\right],
\label{p6} \\
\varpi_cR_{\rm L}\varepsilon\frac{d^2 f}{d\varpi^2}=
-2\sin^2\theta\left[\left(\lambda-\frac{1}{2}\cos\theta\right)\xi_{\varphi}^+
-\left(\lambda+\frac{1}{2}\cos\theta\right)\xi_{\varphi}^-\right],
\label{p7}\\
\varpi_c\frac{d}{d\varpi}\left(\xi_{\theta}^+\gamma^+\right) =
4\lambda\sigma\left(-\frac{\gamma^++\gamma^-}{2\sigma\sin\theta}
-\varpi_c\frac{\cos\theta}{\sin\theta}\frac{d\delta}{d\varpi}
-\sin\theta\xi_r^++\frac{\sin\theta}{x_0}\xi_{\varphi}^+\right),
\label{p8}\\
\varpi_c\frac{d}{d\varpi}\left(\xi_{\theta}^-\gamma^-\right) =
-4\lambda\sigma\left(-\frac{\gamma^++\gamma^-}{2\sigma\sin\theta}
-\varpi_c\frac{\cos\theta}{\sin\theta}\frac{d\delta}{d\varpi}
-\sin\theta\xi_r^-+\frac{\sin\theta}{x_0}\xi_{\varphi}^-\right),
\label{p9}\\
\varpi_c\frac{d}{d\varpi}\gamma^+ =
4\lambda\sigma\left(
-\varpi_c\frac{d\delta}{d\varpi}-\sin\theta\xi_{\theta}^+\right),
\label{p10}\\
\varpi_c\frac{d}{d\varpi}\gamma^- =
-4\lambda\sigma\left(
-\varpi_c\frac{d\delta}{d\varpi}-\sin\theta\xi_{\theta}^-\right),
\label{p11}
\end{eqnarray}
where we neglected the terms $\propto\delta/r$ in (\ref{p8}) and (\ref{p9}).

Comparing the leading terms, we have inside the
layer $\Delta\varpi/R_{\rm L} \sim \lambda^{-1}$
\begin{eqnarray}
\gamma^{\pm} & \sim & h_c^{1/2}\sigma, \\
\xi_{\theta}^{\pm} & \sim  & h_c^{1/4}, \\
\xi_r^{\pm} & \sim & h_c^{1/2}, \\
\Delta\delta & \sim  & h_c^{3/4}/\lambda,
\end{eqnarray}
where $h_c=|h|$. Then the leading terms in
(\ref{p1}) -- (\ref{p5}) for $\Delta\varpi > \lambda^{-1}R_{\rm L}$ are
\begin{eqnarray}
\xi_{\varphi}^+ & = & \frac{1}{x\sin\theta} \approx \frac{1}{x_0},
\label{p12}\\
\xi_{\varphi}^- & = & \frac{1}{x\sin\theta} \approx \frac{1}{x_0},
\label{p13}
\end{eqnarray}
\begin{eqnarray}
2\xi_r^+ & = & (\xi_{\varphi}^+)^2+(\xi_{\theta}^+)^2,
\label{p14}\\
2\xi_r^- & = & (\xi_{\varphi}^-)^2+(\xi_{\theta}^-)^2,
\label{p15}
\end{eqnarray}
\begin{equation}
\zeta=-\frac{(\gamma^++\gamma^-)}{2\sigma\sin\theta},
\label{p16}
\end{equation}
where $x_0=\varpi_c/R_{\rm L}=(2|h|)^{-1/4}$.
Hence, one can totally neglect $\varepsilon f$ and $\delta$ in
(\ref{p1}) -- (\ref{p2}), so Eqns. (\ref{p6}) -- (\ref{p11})
in the region $\Delta\varpi > \lambda^{-1}R_{\rm L}$ can be rewritten as
%
%
\begin{eqnarray}
\varpi_c^2\frac{d^2\delta}{d\varpi^2}=
2\lambda\sin\theta\cos\theta(\xi_{\theta}^+-\xi_{\theta}^-),
\label{p17} \\
\varpi_c\frac{d}{d\varpi}\left(\xi_{\theta}^+\gamma^+\right) =
4\lambda\sigma\left(-\frac{\gamma^++\gamma^-}{2\sigma\sin\theta}
-\varpi_c\frac{\cos\theta}{\sin\theta}\frac{d\delta}{d\varpi}
-\sin\theta\xi_r^+\right),
\label{p18}\\
\varpi_c\frac{d}{d\varpi}\left(\xi_{\theta}^-\gamma^-\right) =
-4\lambda\sigma\left(-\frac{\gamma^++\gamma^-}{2\sigma\sin\theta}
-\varpi_c\frac{\cos\theta}{\sin\theta}\frac{d\delta}{d\varpi}
-\sin\theta\xi_r^
-\right),
\label{p19}\\
\varpi_c\frac{d}{d\varpi}\left(\gamma^+\right)=
-4\lambda\sigma\sin\theta\xi_{\theta}^+,
\label{p20}\\
\varpi_c\frac{d}{d\varpi}\left(\gamma^-\right)=
4\lambda\sigma\sin\theta\xi_{\theta}^-,
\label{p21}
\end{eqnarray}
with all the terms in the right--hand sides of (\ref{p18}) and
(\ref{p19}) being of the same order of magnitude.

As a result, the nonlinear equations (\ref{p17}) -- (\ref{p21}) and
(\ref{p7}) give the following simple asymptotic solution
\begin{eqnarray}
\gamma^{\pm} & = & 4\sin^2\theta\sigma(\lambda l)^2, \\
\xi_{\theta}^{\pm} & = & \mp 2\sin\theta\lambda l, \\
\Delta\delta & =  &
-\frac{4}{3}\sin^2\theta\cos\theta\lambda^{-1}(\lambda l)^3,
\label{o2}\\
\Delta(\varepsilon f) & = & \sin^2\theta\cos\theta\lambda^{-2}(\lambda l)^2,
\label{o3} \\
\zeta & = & -4\sin\theta(\lambda l)^2,
\label{o4}
\end{eqnarray}
where now $l=\Delta\varpi/\varpi_c$. It is important that the last expressions
are correct for arbitrary relations between $\gamma_{\rm in}^3$ and $\sigma$.
As we can see, in the narrow layer $\Delta\varpi = \varpi_c/\lambda$
the particle energy increases up to the value $\sim \sigma$ which
corresponds to the full conversion of the electromagnetic
energy to the energy of particles. For this reason we have
here $|\zeta| \sim 1$, which just means the diminishing of the
toroidal magnetic field determining the flux of the electromagnetic
energy.  On the other hand, the variation of the electric potential
remains small $\delta \sim \lambda^{-1}$, to say nothing about
the variation of the magnetic surfaces $\Delta\varepsilon f \sim
\lambda^{-2}$. These results coincide exactly with our previous
evaluations (Beskin et al 1983) allowing us to neglect
variations of the electric potential and the poloidal magnetic
structure in the 1D cylindrical case.

The latter result has a simple physical explanation.
Indeed, the diminishing of the toroidal magnetic field is connected
with the $\theta$--component of the electric current which
is produced by all the particles moving in opposite directions.
On the other hand, the change of the electric potential
depends on the small difference between the electron and positron
densities.  As a result, according to (\ref{o3}) and (\ref{o4}),
the change of the toroidal magnetic field is just $\lambda$ times
larger than the change of the electric potential.  Unfortunately,
it is impossible to consider this region more thoroughly because
for $\lambda l \sim 1$ we have $\xi_{\theta}^{\pm} \sim 1$ and
$\xi_{r}^{\pm} \sim 1$, i.e. the linear approximation
(\ref{b1}) -- (\ref{b2}) itself becomes incorrect.

It is necessary to stress as well that we do not include into consideration
the radiation reaction force
\begin{equation}
F_x^{\rm (rad)} = -\frac{2}{3}\frac{e^4}{m^2c^4}\gamma^2
\left[(E_y - B_z)^2 + (E_z - B_y)^2\right],
\label{frad}
\end{equation}
which can be important for large enough particle energy.
Comparing (\ref{frad}) with appropriate terms in
(\ref{p18}) -- (\ref{p21}) one can conclude that the radiation
force can be neglected for $\sigma < \sigma_{\rm cr}$, where
\begin{equation}
\sigma_{\rm cr} = \left(\frac{c}{\lambda r_e \Omega}\right)^{1/3}
\approx 10^6,
\end{equation}
and $r_e = e^2/mc^2$ -- classical electron radius.
This relation can be rewritten in the form
\begin{equation}
\frac{\Omega R}{c} < 3 \times 10^{-3}B_{12}^{-3/7} \lambda_4^{2/7}
\end{equation}
which gives
\begin{equation}
P > 0.06B_{12}^{3/7} \lambda_4^{-2/7} s.
\end{equation}
Hence, for most radio pulsars the radiation force indeed can be neglected.
As to the pulsars with $\sigma > \sigma_{\rm cr}$,
it is clear that for $\gamma > \sigma_{\rm cr}$ the radiation force
becomes larger than the electromagnetic one and strongly inhibits
any further acceleration.
As a result, we can evaluate the maximum gamma--factor which can
be reached during the acceleration as
\begin{equation}
\gamma_{\rm max} \approx \sigma_{\rm cr} \approx 10^6.
\end{equation}

\section{Discussion}

Thus, on a simple example it was demonstrated that for real
physical parameters of the magnetosphere of radio pulsars
($\sigma \gg 1$ and $\lambda \gg 1$)
the one--fluid MHD approximation remains true in the whole region
within the light surface $|{\bmath E}| = |{\bmath B}|$.
On the other hand, it was shown that in a more realistic 2D case
the main properties of the boundary layer near the light surface
existing for small enough longitudinal currents $I < I_{GJ}$
(effective energy transformation from electromagnetic field to particles,
current closure in this region, smallness of the disturbance of
electric potential and poloidal magnetic field)
remain the same as in the 1D case considered previously (Beskin et
al 1983).

It is necessary to stress the main astrophysical consequences of our
results. First of all, the presence of such a boundary layer explains
the effective energy transformation of electromagnetic energy
into the energy of particles. As was already stressed, now the
existence of such an acceleration is confirmed by observations of
close binaries containing radio pulsars (as to the particle
acceleration far from a neutron star, see e.g. Kennel \& Coroniti
1984, Hoshino et al 1992, Gallant \& Arons 1994). Simultaneously,
it allows us to understand the current closure in the pulsar
magnetosphere.
 Finally, particle acceleration results in the
additional mechanism of high--energy radiation from the boundary of
the magnetosphere (for more details see Beskin et al 1993).

Nevertheless, it is clear that the results obtained do not solve
the whole  pulsar wind problem. Indeed, as in the cylindrical
case, it is impossible to describe the particle motion outside
the light surface. The point is that, as one can see directly from
Eqn. (\ref{o2}), for a complete conversion of electromagnetic
energy into the energy of particles it is enough for them
to pass only $\lambda^{-1}$ of the total potential drop between
pulsar magnetosphere and infinity. It means that
the electron--positron wind propagating
to infinity has to pass the potential drop which is much larger
than their energy. It is possible only in the presence of
electromagnetic waves even in an axisymmetric magnetosphere which
is stationary near the origin. Clearly, such a flow cannot be
considered even within the two--fluid approximation.  In our opinion,
it is only a numerical consideration that can solve the problem
completely and determine, in particular, the energy spectrum of
particles and the structure of the pulsar wind.  Unfortunately, up
to now such numerical calculations are absent.

\section*{Acknowledgments}

The authors are grateful to I.~Okamoto and H.~Sol for fruitful discussions.
VSB thanks National Astronomical Observatory, Japan for hospitality.
This work was supported by INTAS Grant 96--154 and by Russian Foundation
for Basic Research (Grant 96--02--18203).
%

\end{document}